\documentclass{article}

\usepackage{graphicx}  
\usepackage{amsmath}   
\usepackage[compress]{cite}
\usepackage{amssymb}   

\usepackage[active]{srcltx}

\usepackage{color}

\def\eq#1{{Eq.~(\ref{#1})}}

\def\cc{{cosmological\ constant}}

\def\md{microscopic degrees of freedom of the spacetime}

\title{Do We \textit{really} Understand the Cosmos?}

\author{T. Padmanabhan\\
IUCAA, Pune University Campus, \\
Ganeshkhind, Pune 411007, India.\\
email: paddy@iucaa.in
}

\date{ }

\begin{document}

\maketitle

\begin{abstract}
Our \textit{knowledge} about the universe has increased tremendously in the last three decades or so --- thanks to the 
progress in observations --- but our \textit{understanding} has improved very little. There are several fundamental questions about our universe for which we have no answers  within the current, operationally very successful, approach to cosmology.  Worse still, we do not even know how to address some of these issues within the  conventional approach to cosmology. This fact suggests that we are  missing some important theoretical ingredients in the overall description of the cosmos. 
I will argue that these issues --- some  of which are  not fully appreciated or emphasized in the literature --- demand a paradigm shift: We should not think of the universe as described by a specific solution to the gravitational field equations; instead, it should be treated as a special physical system governed by a different mathematical description, rooted in the quantum description of spacetime. I will outline how this can possibly be done.
\end{abstract}

\tableofcontents

\section{Motivation}

Spectacular progress in cosmological observations in the last four decades has helped us to develop a standard model of the universe which is very successful. In this model, the smooth universe is described by a \textit{specific} solution to the field equations of gravity, say, Einstein's equations, and can be parameterized by a small set of numbers ($H_0,\Omega_B,\Omega_{\rm DM}, \Omega_{\rm DE},\Omega_R$ ...with their usual meanings). In  addition, the formation of structures in the universe is described quite adequately in terms of the growth of perturbations around this smooth universe. These perturbations, generated during an inflationary phase\footnote{I would love to have a viable alternative to the inflationary generation of perturbations, but there is none which can be considered a worthy challenger. So, in this article I will accept the inflationary paradigm as a working hypothesis.}, can be characterized by a power spectrum $P(k) = Ak^n$ with two more parameters, $A$ and $n$. Both theory and observations are  mature enough today to test not only the lowest order predictions (for e.g., scale invariance of the perturbations, corresponding to $n=1$), but also higher order effects (like, for e.g. the deviation $(n-1)$, in specific models). Thus, on the whole, the description of the universe can be based on  a set of well defined  parameters which are directly observable.

At the next level of probing, such a description encounters three kinds of difficulties, of which the first two are well-known in the literature and the third one will be the core topic of discussion in this article. 

The first kind of difficulty is related to technical issues and details in the model. The following questions, for example, belong to this set: Can we correctly describe the properties and statistics of dwarf galaxies? Do we understand the detailed mechanism which caused the reionization in the universe? Most cosmologists (including me) believe that it is just a question of time before we have satisfactory and consistent answers to such issues within the standard description. 

The second kind of difficulty which arises in cosmology is related to the description of the matter sector. Examples are questions like: What is the nature and abundance of the dark matter\footnote{Verification of Einstein's equations at cosmological scales require testing the hypothesis $G^a_b-\kappa T^a_b=0$ where $\kappa=8\pi G$. When the directly observed values of these two tensors $G^a_b(obs)$ and $T^a_b(obs)$ lead to $G^a_b(obs)-\kappa T^a_b(obs)\equiv \kappa Q^a_b\neq0$, as it happens in our universe,  Einstein's theory appears to flunk the test. We can then either postulate a modified  matter tensor $T^a_b=T^a_b(obs)+Q^a_b$, (as done in the case of dark matter) or a modification of theory by $G^a_b=G^a_b(obs)-\kappa Q^a_b$, (as done in the case of dark energy which I take to be the  cosmological constant). I will accept both these modifications, viz, the postulates of dark matter and the cosmological constant, in this article. One can question these assumptions, but again I find that all alternatives are much worse theoretically.} particle? How can we explain the baryon-to-photon ratio in our universe? These issues are more fundamental than the first kind of problems but most of us believe that we do have an algorithmic procedure available to attack these  problems, \textit{within the framework of conventional cosmology}. For example, a successful extension of the standard model in high energy physics might allow us to compute such numbers from first principles. The current difficulty is then only due to our inadequate understanding of particle physics at high enough energies. 

The third kind of problems --- which, as I said, we will be concerned with --- are those which we have no clue as to how to address. The most important example in this category is the extremely tiny --- but non-zero --- value of the cosmological constant.\footnote{In this article, I will assume that dark energy \textit{is} cosmological constant. Other explanations for dark energy are more ad-hoc, not demanded by observations, do not explain why cosmological constant is zero and  leaves the fine tuning problem unanswered. I do not think these models are better alternatives to the postulate of a cosmological constant.} As regards this set, I am not so much concerned about the lack of a viable solution as with the fact that we do not even know how to properly attack these problems within the framework of conventional cosmology. In some cases, which I will discuss, it is not even clear how to precisely state these problems within the context of the standard model of cosmology.

After some clarifications on the notion of expansion of the universe (Sec. \ref{sec:intro}) I will describe, in Sections \ref{sec:allok} to \ref{sec:timearrow}, these foundational conundrums in cosmology. Based on this discussion, I will argue (see Sec. \ref{sec:altpara}) that it is fundamentally incorrect to describe the universe as a specific solution to the gravitational field equations. Instead we should think of the universe a special system and look for a  different paradigm to describe its evolution. I will suggest, towards the end of the article, some possible ingredients of such a paradigm and explain (see Sec. \ref{sec:cosmin}) how it can solve the cosmological constant problem. I will use the mostly positive signature and set $c=1, \hbar=1$ and (occasionally) $G=1$. The Greek indices range over 1,2,3 while the Latin indices range over 0-3.

\section{Expansion of the Universe is in the eye of the beholder}\label{sec:intro}

The standard cosmological model is based on a specific solution to gravitational field equations. It is generally believed that one key feature of this solution is the `expansion' of the smooth background universe which is supposed to distinguish the Friedmann solution from, say, the spacetime describing the region around the Sun. What is not adequately emphasized is the fact that the standard notion of expansion depends on the coordinates you choose to describe the Friedmann model. 
Geodesic observers in a spacetime will interpret not only the Friedmann metric but also the Schwarzschild metric as `expanding', while non-geodesic observers can find both of them non-expanding --- in a precise sense, described below. This tells you that one need not associate the theoretical difficulties we will be discussing later too strongly with the notion of expansion. Since this result, unfortunately, has not attracted necessary attention in the literature\footnote{I have discussed a few of these issues in Ref.\cite{oxcamlect1} as well as in Ref.\cite{hptp}.} and comes as a surprise even to some experts,  I will introduce the Friedmann model from a different perspective and highlight this fact. 

It seems reasonable to assume that the smooth, large scale, spatial 3-geometry of the universe should be homogeneous and isotropic and hence must have a constant 3-curvature which can be taken to be $k=0,-1$ or $+1$. I will confine my attention to models in which $k=0$, so that the spatial sections are flat.\footnote{This is certainly preferred by both observations and theory. Most of the discussion can be generalized to $k\neq0$, but not all.} Such a maximally symmetric 3-space should also necessarily be spherically symmetric about any given spatial origin. It is, therefore, natural to foliate the 3-space by 2-dimensional spherical surfaces with the metric $dl^2 = r^2 (d\theta^2+\sin^2\theta d\phi^2)\equiv r^2d\Omega^2$ where $r$ is a radial coordinate with a clear physical meaning: $r=[A/(4\pi)]^{1/2}$ with $A$ being the \textit{proper}  area of the foliating 2-surfaces. Consider now a spacetime metric given by 
\begin{equation}
 ds^2 = - \frac{1}{24\pi G} \frac{d\rho^2}{\rho (\rho +p)^2} + \left[ dr + \frac{r}{3(\rho+p)} d\rho \right]^2 + r^2 d\Omega^2
 \label{frw1}
\end{equation} 
where the coordinates are chosen to be $(\rho,r,\theta,\phi)$ and $p = p(\rho)$ is a specified function. 
This metric describes our universe with $p=p(\rho)$ representing an effective equation of state for the matter with $p$ and $\rho$ interpreted as total pressure and total density! 
If you compute the $G^a_b$ for this metric, you will find that it satisfies Einstein's equations $G^a_b=\kappa T^a_b$ with a source energy momentum tensor $T^a_b=[\rho+p(\rho)]u^au_b+p(\rho)\delta^a_b$ where $u^a$ is the four-velocity of \textit{geodesic} observers in the spacetime.  This, in turn implies that
$\rho=T_{ab}u^au^b,p=(1/3)T_{ab}h^{ab}$ where $h_{ab}=g_{ab}+u_au_b$ is the projection tensor orthogonal to the four-velocity $u^a$ of the geodesic observers and $T_{ab}$ is essentially determined by $\rho$ and the function $p(\rho)$.

The line element in \eq{frw1} is remarkable in the sense that the spacetime geometry could be expressed directly in terms of the variables which occur in the matter sector of the theory through $T^a_b$. That is, we have now solved the Einstein's equations $G^a_b = \kappa T^a_b$ for the metric\footnote{The metric, as it is written,  has a singularity if we choose the equation of state to be exactly $p=-\rho$; but this can be handled by a careful limiting procedure or in a different coordinate system. Our universe is \textit{never} described by a strictly $p=-\rho$ equation of state. Both during the inflationary phase as well as the late time acceleration, this equation of state is approached only asymptotically, and hence the line element is well-defined.} and have expressed the components of  the metric $g_{ab}$ directly in terms of the components of the stress tensor $T^a_b$. This is, of course, impossible to do in general but works in the case of the universe only because of its high level of symmetry.

To reduce the metric in \eq{frw1} into the conventional form is quite straightforward. Given a function $p=p(\rho)$, compute the indefinite integral
\begin{equation}
  t = - \left( \frac{1}{24\pi G} \right)^{1/2} \int \frac{d\rho}{(\rho+p)\sqrt{\rho}}
 \label{eqn4.2}
\end{equation} 
to obtain the function $t=t(\rho)$. Invert this function, locally, to determine $\rho = \rho(t)$ and thus $p = p(\rho) = p(t)$, obtaining $\rho$ and $p$ as functions of $t$.  Define, for convenience, the function $H(t)$ through 
\begin{equation}
H(t) \equiv - \frac{1}{3} \frac{d\rho}{dt} \frac{1}{(\rho+p)}  = \left( \frac{8\pi G\rho}{3}\right)^{1/2}
 \label{eqn4.3}
 \end{equation} 
 where the second equality follows from \eq{eqn4.2}. 
  Transform from the coordinates $(\rho, r, \theta, \phi)$ to the coordinates $(t,r,\theta,\phi)$ and\footnote{When $\rho$ is a monotonic function of $t$ you can switch from $\rho$ to $t$ trivially; if not, you can still do it locally and glue the definitions together appropriately.}  you will find that the line interval in \eq{frw1}
 becomes
 \begin{equation}
 ds^2 = - dt^2 + \left( dr - H(t) r dt\right)^2 + r^2 d\Omega^2
 \label{eqn4.4}
 \end{equation} 
Some of you will recognize this line element as representing the Friedmann model in the Painleve type coordinates; if you don't, introduce a function $a(t)$ and a coordinate $x$ through the relations 
\begin{equation}
 H(t)\equiv \frac{\dot a}{a}; \qquad x \equiv \frac{r}{a}
 \label{eqn4.5}
 \end{equation} 
 and you will find that the line interval in the coordinates $(t,x,\theta,\phi)$ is given in the familiar form:
 \begin{equation}
  ds^2 = - dt^2 + a^2(t) \left( dx^2 + x^2 d\Omega^2\right)
 \label{eqn4.6}
 \end{equation} 

The line element in \eq{eqn4.4} contains a single unknown function of time, $H(t)$. The metric as well as the field equation can be expressed entirely in terms of the function $H(t)$. But  the function $a(t)$, \textit{defined} through the first equation in \eq{eqn4.5}, is not unique and has a scaling degree of freedom, $a\to \lambda  a$. This is obvious when the metric is written as in \eq{eqn4.4} because $H$ is invariant under constant rescaling of $a(t)$. This is not apparent if we start with the standard form of the Friedmann metric in \eq{eqn4.6} unless we also rescale $x$. This is usually considered to be a rather trivial matter but it is not. 
Equation (\ref{eqn4.5}) clearly shows that for a given $H(t)$ determined by the source, the corresponding $a(t)$ is \textit{not} unique and is arbitrary with respect to a scaling by a constant. Such a scaling freedom does \textit{not} exist if we use the coordinates in \eq{eqn4.4} or \eq{frw1} to describe the Friedmann geometry. If we rescale $a$, then the second equation in \eq{eqn4.5} tells us that $x$ is \textit{automatically} rescaled, leaving $r$ fixed. I stress that $r$ has a direct geometrical meaning $(A/4\pi)^{1/2}$ in terms of the area of the $t=$ constant, $r=$ constant surface. So, the real origin of the scaling freedom in $a(t)$ is from the fact that the geometrical description in \eq{eqn4.4} or \eq{frw1} cares only for $H(t)$ and that the $a(t)$ arises through the definition in the first equation in \eq{eqn4.5}. 
  This, in turn, implies that we have the freedom to set $a(t) =1$ for some value of $t$ while describing the universe. I will say more about this choice later on.

 The line elements in \eq{frw1} or \eq{eqn4.2} give us a very different pictures about the `expansion of the universe', compared to the line element in \eq{eqn4.6}. Observers comoving with the coordinates $(t,r,\theta,\phi)$ or $(\rho,r,\theta,\phi)$ --- i.e., observers with world lines having $r,\theta,\phi$ fixed --- will find that the spatial cross-sections [corresponding to $t=$ constant or $\rho=$ constant] are described by \textit{flat Euclidean 3-space}. In particular, the volume enclosed by the $r=$ constant surface in 3-space is just $(4\pi/3)r^3$ and the area of the $r=$ constant surface is $4\pi r^2$. Neither the volume nor the area ``expands'' as time evolves in this coordinate system! In contrast, observers using the coordinate system $(t,x,\theta,\phi)$ will find that the spatial cross-sections are Euclidean 3-spaces  scaled by an overall time dependent factor $a(t)$. The volume enclosed by the surface $x=$ constant is $(4\pi/3) x^3 [a(t)]^3$ and the area of the $x=$ constant surface is $4\pi x^2[a(t)]^2$. Both this volume and the area change with time and the universe ``expands'' in this coordinate system if $a(t)$ is an increasing function of time. 
 
The above result demonstrates the title of this subsection.
The observers with $\mathbf{x}=$ constant are geodesic observers and the clocks carried by them measure the cosmic time $t$. These observers see the universe as expanding. The observers following the world line $\mathbf{r}=$ constant 
are \textit{not} geodesic observers. When we use the metric in \eq{eqn4.4}, the geodesics are described by the equation 
\begin{equation}
  r \exp \left[-\int H(t)dt\right] = \text{constant}
 \label{eqn1.64}
\end{equation} 
Since we like to think of galaxies to be in geodesic motion in our universe (though no real galaxy follows the geodesic of the smooth Friedmann model), it is useful to use coordinates in which each galaxy has a constant value of
$x,\theta,\phi$ rather than a constant value of $r,\theta,\phi$. This is merely a question of convenience and is not fundamental. 
 The geometrical features of our universe (like, e.g. redshift) do not change under coordinate transformations, and we can indeed talk of all physical phenomena using the metric in \eq{eqn4.4} without ever introducing the notion of an `expanding' universe.
 
 In case you find this surprising, let me assure you that there is no swindle. The ``expansion'' of a spacetime defined in terms of the increase of, say, the proper areas of the surfaces with $t=$ constant, $r=$ constant,  is \textit{always} a coordinate dependent effect and can occur in \textit{several} spacetimes. Consider, for example, the following metric:
 \begin{equation}
ds^2 = - c^2 dt^2 + \frac{4}{9} \left[ \frac{9GM}{2(x+ct)}\right]^{2/3} dx^2 + \left[ \frac{9GM}{2} ( x + ct)^2 \right]^{2/3} 
[d\theta^2+\sin^2\theta d\phi^2]
  \label{eqn7.52}
 \end{equation} 
 The proper area $A$ of the 2-surfaces with $t=$ constant, $x=$ constant,  increases with time as $A\propto (x+ct)^{4/3}$; similarly, the volume enclosed by the surface
 $t=$ constant, $x=$ constant,  also increases with time. The observers using $t, x, \theta, \phi$ in the spacetime described by the metric in \eq{eqn7.52} can claim --- just like the observers using the coordinates in \eq{eqn4.6} --- that their spacetime is expanding.\footnote{More formally, one can introduce in any spacetime the notion of a congruence of \textit{geodesic} observers with a geodesic velocity field $u^i(x)$. You can define an expansion of this congruence by $\theta\equiv \nabla_i u^i$ which appears to give a  geometric, coordinate-independent, definition of expansion in the Friedmann universe. This is true but $\theta$ will be non-zero for the geodesic congruence in most spacetimes, including the spacetime around the Sun. If you introduce synchronous coordinates in which the metric is $ds^2=-dt^2+h_{\alpha\beta}dx^\alpha dx^\beta$, then the geodesic velocity field is $u^i=\delta^i_0$ and $\theta=\partial_t[\ln\sqrt{h}]$ will be non-zero in general. In the Friedmann universe, the maximal symmetry of space itself gives you a preferred timelike vector which coincides with the velocity vector of geodesic observers; so  $\theta$ is independent of $x^\alpha$, which will not happen in general. The definition of expansion still remains linked to a choice of observers in one way or another even when you use such --- more geometric --- ideas.}
 But the metric in \eq{eqn7.52} describes the spacetime outside a spherical star like the Sun! Instead of the standard Schwarzschild coordinates, we are using the coordinates appropriate to freely falling observers \cite{tpgrav}. Just as in the case of  \eq{eqn4.6}, the time coordinate $t$ in \eq{eqn7.52} denotes the proper time shown by the geodesic observers with trajectories $\mathbf{x} = $ constant. So, even a static spacetime can appear to be expanding when you use geodesic coordinates. The co-moving observers in the standard Schwarzschild coordinates, of course, are non-geodesic observers just as the observers who use \eq{frw1} or \eq{eqn4.4}. 
 There is nothing sacred about geodesic observers and, in fact, we hardly use the coordinates adapted to geodesic observers anywhere in GR except in the case of cosmology. 
 
 The reason I brought in the coordinate dependence of the notion of ``expanding'' universe is the following: As we proceed to discuss several conundrums in Friedmann model,  they may appear to be  related to the fact that spatial sections of the universe are expanding. This, however, is incorrect.  One can discuss all of cosmology including the big bang singularity, the inflationary phase etc.  without this notion, by choosing  a different set of coordinates. The metric will still be time dependent --- through $H(t)$ in \eq{eqn4.4} and through $p(\rho)$ in \eq{frw1} --- but the spatial sections will be non-evolving Euclidean 3-space. So it is  conceptually inappropriate to attribute all the theoretical issues which we will come across to the fact that the universe is expanding. 
 
 \section{Everything is allowed in cosmology}\label{sec:allok}
 
 In general relativity, we think of the field equation $G^a_b = \kappa T^a_b$ as determining the metric tensor $g_{ab}$ for a \textit{given} source described by $T^a_b$. We never ask where $T^a_b$ came from; nor do we  impose any constraints on the nature of $T^a_b$ except to demand that $\nabla_a T^a_b =0$. This works quite well in all non-cosmological contexts like e.g., when you want to determine the gravitational field around a binary  pulsar or the gravitational field produced by a massive rotating black hole. You choose the appropriate $T^a_b$, solve Einstein's equation to get $g_{ab}$ and then work out all the properties. 
 
 This approach runs into a curious problem in the context of cosmology. If you use the standard Friedmann coordinates in \eq{eqn4.6}, then the energy momentum tensor should necessarily be of the form $T^a_b = $  dia $(-\rho(t),p(t),p(t),p(t))$ having two undetermined functions of time. The metric has one undetermined function $H(t)$ and these \textit{three} functions are related by \textit{two} independent components of Einstein's equations: 
\begin{equation}
 \rho(t)=\frac{3H^2}{8\pi G}; \quad p(t)=- \frac{1}{8\pi G}[3H^2+2\dot H]
 \label{eqn4.7}
 \end{equation} 
  This, in turn,  tells you two things: First, the Friedmann model of the universe is under-determined unless you externally specify a relation between $p$ and $\rho$ and put some conditions on what kind of $p$ and $\rho$ are acceptable.  If you don't (and most modern cosmologists don't)  you can have any evolutionary model of the universe described by any real function $a(t)$ [or $H(t)$] by a suitable choice of the equation of state $p=p(\rho)$! All you need to do is to choose your favourite model of evolution described by some $a(t)$ [which could give you a universe with any kind of features you like], compute $\rho(t)$ and $p(t)$ through \eq{eqn4.7} and eliminate $t$ between these two variables to determine an equation of state $p=p(\rho)$ for the material medium populating the universe.\footnote{If the universe is populated by several species of energy density, then $\rho$ and $p$ will be the sum of densities and pressures of each species and, of course, each of them will have their own equation of state. But since total pressure $p(t)$ and total density $\rho(t)$ are just functions of time, you can eliminate $t$ between the two and determine an effective equation of state $p=p(\rho)$ for an `effective fluid' which will produce the same geometry. You can also play the same game --- as is often done in various disguises --- using a scalar field with a potential $V(\phi)$. I have given an explicit recipe for constructing a $V(\phi)$ for \textit{any} $a(t)$ in Ref.\cite{acctach}.} The equation of state will be weird and  fine-tuned but such models are routinely published in the literature. Somewhat gratifyingly, $\rho(t)$ given by \eq{eqn4.7} will be positive definite but $p$ could have either sign. A source with $p<0$ would have been unthinkable some five decades back but today, negative pressure sources are not only considered acceptable but also respectable by the current generation of cosmologists. So you can publish a paper with any evolutionary history $a(t)$ if you do not care for laboratory justification  for the energy momentum tensor. \textit{Cosmological evolution is, in principle, fundamentally unconstrained} which is an issue we need to recognize.
  
  The situation is made worse by the following factors: If you start with an equation of state which \textit{is} tested and justified in the laboratory, you will most probably have $\rho>0, p>0$. When you evolve the universe backward in time, you will eventually reach energy scales which are not tested in the laboratory. This is going to happen irrespective of how high an energy scale you can explore in the laboratory. So, unless you discover in the lab, forms of matter with an equation of state which violates $p+\rho >0$ and/or $\rho+3p>0$, you are always going to hit an unknown domain. You cannot do cosmology by specifying a lab-tested $T^a_b$ and solving for $g_{ab}$ if (i) your lab-tested $T^a_b$ has $\rho>0$ and $p>0$ and (ii) you go sufficiently far into the past. This makes cosmology rather special from the point of view of solving Einstein's equation. You \textit{need to} postulate different forms of $T^a_b$, work out the observational consequences and iterate on the procedure. Unfortunately this never works out satisfactorily in practice. The near-infinite number of inflationary models available in the market is a simple proof that this procedure is unlikely to give you a predictive, unique description of the very early phases of the universe. 
  
  There is also a deeper conceptual issue. It was known right from the early days of GR that the equation $G^a_b = \kappa T^a_b $ equates something extraordinarily beautiful on the left hand side to some ugly structure on the right hand side. This distinction gets worse as we start probing the earlier and earlier phases of the universe. For example, most of the scalar field potentials used in inflationary model building have no particle physics justification and are theoretically unacceptable from the point of view of quantum field theory --- even as an effective field theory. Add to it the fact that both matter and geometry has to ``come out of nothing''  at the beginning and you can see the knot we have tied ourselves in. This is the first of a series of questions I want to raise for which we not only don't have a solution but --- even worse --- we do not even know how to approach the problem.
  This difficulty arises because \textit{the geometry does not constrain the matter sector sufficiently} in standard GR but leaves us with far too many choices.

  \section{The strange (and stranger) numbers which characterize our universe}
  
  Observations are consistent with the idea that our universe can be described in terms of three distinct evolutionary phases: (i) An inflationary phase with an equation of state $p\approx -\rho$; (ii) a radiation dominated phase with the equation of state $p\approx (1/3) \rho$ followed by a matter dominated phase with $p\approx 0$. (iii) A late time accelerated phase with the equation of state $p\approx-\rho$ which I will take to be dominated by the cosmological constant. The standard way of describing the evolutionary history of such a model is through the equation 
  \begin{equation}
H^2(t) = \frac{\dot a^2}{a^2} =
\begin{cases}
H_{\rm inf}^2\qquad (a<a_{\rm rh})\\
\noalign{\medskip}
 H_0^2 \left[ ( 1 - \Omega_R - \Omega_m)+ \Omega_R a_0^4/a^4 + \Omega_m a_0^3/a^3\right]\qquad (a>a_{\rm rh})
\end{cases}
   \label{eqn4.8}
  \end{equation} 
where $a_{\rm rh}$ is the epoch of reheating at which the inflation ended. (I have assumed instant reheating  and set $k=0$ for simplicity).  Such a description uses the constant parameters,  $H_{\rm inf}, H_0$, $\Omega_R$, $\Omega_m$ and $a_0$. (We usually set to $a_0=1$ but there is a subtlety about this choice which I will come back to). 

While these parameters are very convenient to compare observations with theory, they are completely unsuited for describing the universe as a physical system. For example, cosmologists living in a star system located in a galaxy at $z=8$ will use corresponding parameters evaluated at $z=8$ which, of course, will differ numerically from the ones we use. In other words, the parameters used in \eq{eqn4.8} have no epoch invariant significance and are tied to a very special epoch at which the CMB temperature is 2.73 K. This is unsatisfactory when we want to think of the universe as a physical system described by certain cosmic constants. It is necessary to describe the evolution of the universe using constant parameters which will have epoch-independent significance.
  
  This  is fairly easy to do and, in fact, one can do it in an infinite number of ways. One convenient set of parameters to use are the following: (i) We describe the inflationary phase by a constant density $\rho_{\rm inf}$ with the equation $H^2(t)\approx (8\pi G\rho_{\rm inf})/3$ which will replace the first part of \eq{eqn4.8}. (ii) Similarly, we introduce another constant density $\rho_\Lambda $ corresponding to the cosmological constant so that at very late times the universe will be described by the deSitter expansion with  $H^2(t) \approx (8\pi G\rho_\Lambda)/3$. (iii) To describe the radiation and matter dominated phase, it is convenient to introduce another \textit{constant} density 
  \begin{equation}
  \rho_{\rm eq}\equiv\frac{\rho_m^4(a)}{\rho_R^3(a)}=\sigma T_{eq}^4
   \label{eqn4.9}
  \end{equation}
  where $\sigma$ is the Stefan-Boltzmann constant and the second equality defines the temperature $T_{\rm eq}$.  We can also introduce the parameter $a_{\rm eq} $ by the epoch-independent definition 
  \begin{equation}
 a_{\rm eq}\equiv \frac{a \rho_R(a)}{\rho_m(a)}
   \label{eqn4.10}
  \end{equation}
  and work with the variable $x\equiv (a/a_{\rm eq})$.  Equation (\ref{eqn4.8}) can now be rewritten in the form
  \begin{equation}
  \left(\frac{\dot x}{x}\right)^2=
  \begin{cases}
  &(8\pi G/3)\ \rho_{\rm inf}\\
    & (8\pi G/3)\ \left[ \rho_\Lambda + \rho_{\rm eq} \left( 
 x^{-3} + x^{-4}\right)\right]
 \end{cases}
   \label{eqn4.11}
  \end{equation}
  in terms of the three densities $(\rho_{\rm inf}, \rho_{\rm eq}, \rho_\Lambda)$. This is a much more meaningful way of describing our universe than by using the parameterization in \eq{eqn4.8}. In particular, our cosmologist friend who lived in the $z=8$ galaxy would have written exactly the same equation with exactly the same numerical values\footnote{More precisely, if you divide each of these densities by the Planck density $\rho_{\rm Pl} = c^5/G^2\hbar$, you will get three dimensionless numbers which will be the same as those used by the $z=8$ cosmologist. So it does not matter that we are using the CGS system which might not have existed at $z=8$!} for $(\rho_{\rm inf}, \rho_{\rm eq}, \rho_\Lambda)$. 
  We could also convey to the $z=8$ cosmologist our normalization convention for $a(t)$: 
  We tell her to set $a(t)=1$ at the epoch when the CMB temperature was equal to $T_{\rm eq}$; this is again an invariant statement characterizing the description of our universe.\footnote{This is  a good time to point out the fact that there are certain constants in the universe the numerical value of which we cannot determine uniquely. For example, consider  the combinations like $aT(a)$ or $a\rho_R(a)/\rho_M(a)$. These quantities remain constant, independent of the epoch $a$ at which they are measured, as the universe evolves. But their numerical value depends on the numerical value you attribute to $a_0$ which is not determined by theory. Recall that Einstein's equation only fixes $H(t)$ and not $a(t)$. So while we know that $aT(a) = a_0T_0 = a_{\rm eq} T_{\rm eq}$, we will never know its numerical value without making an additional assumption.  This is why it is convenient to use the scaling freedom and set $a=1$ at the epoch when the radiation temperature was equal to $T_{\rm eq}$. This is a normalization which is independent of the current epoch and something with which our cosmologist at $z=8$ will agree.}
  In other words, \eq{eqn4.11} describes the universe as a physical system (like, for e.g. an elastic solid) determined by certain constants (like, for e.g. the Young's modulus  etc. for a solid). 
  
  Obviously it is meaningful to ask why our universe has certain numerical values for $(\rho_{\rm inf}, \rho_{\rm eq}, \rho_\Lambda)$ in terms of, say, the Planck density. From observations we know  that: 
  \begin{equation}
 \rho_{\rm inf}<(1.94\times 10^{16}\ \text{GeV})^4
   \label{eqn4.12}
  \end{equation}
  \begin{equation}
  \rho_{\rm eq}=  \frac{\rho_m^4}{\rho_R^3}
=[(0.86\pm 0.09) \ \text{eV} ]^4
   \label{eqn4.13}
  \end{equation}
  \begin{equation}
  \rho_\Lambda=[(2.26\pm 0.05)\times 10^{-3}\text{eV}]^4
   \label{eqn4.14}
  \end{equation}
  in natural units with $c=1,\hbar=1$. Several comments are in order vis-a-vis these values. 
  
  Today we have no firm theory which determines the numerical values of any of these three densities. However, the \textit{nature} of our ignorance about them differs   significantly. The bound in \eq{eqn4.12} could actually be replaced by a numerical value if future observations determine the energy scale of inflation. Further, high energy physics does provide a glimmer of hope in eventually coming up with some sensible\footnote{By `sensible', I mean a model of inflation in which the scalar field driving the inflation, for example, serves some useful purpose other than just driving the inflation.} model of inflation which will determine the density in \eq{eqn4.12}. 
  
  Consider next the numerical value of $\rho_{\rm eq}$. 
  From the definition, we can relate $\rho_{\rm eq}$ to the ratio between the number density of the photons and the number density of  matter particles:
\begin{equation}
\rho_{\rm eq}=  \frac{\rho_m^4}{\rho_R^3}
=C \frac{(n_{\rm DM} m_{\rm DM}+n_{\rm B} m_{\rm B})^4}{n_\gamma^4}
=C\left[m_{\rm DM} \left(\frac{n_{\rm DM}}{n_\gamma}\right) + m_{\rm B} \left(\frac{n_{\rm B}}{n_\gamma}\right)\right]^{4}
\label{heprhoeq}
\end{equation} 
where $C =  15^3 (2 \zeta(3))^4  c^3 /\pi^{14} \hbar^3$   is a numerical constant, $n_{\rm DM},n_{\rm B},n_\gamma$ are the current number densities of dark matter particles, baryons and photons respectively and $m_{\rm DM},m_{\rm B}$ are the masses of the dark matter particle and baryon.
 We expect the physics at (possibly) GUTs scale to determine the ratios $(n_{\rm DM}/n_\gamma)$ and $(n_{\rm B}/n_\gamma)$ and specify $m_{\rm DM}$ and $m_{\rm B}$.
 Indeed, we have a framework to calculate these numbers in different models of high energy  physics (for a review, see e.g., \cite{mazumdar}) though none of these models can be considered  as compelling at present.

  Thus we do have a possible theoretical framework for determining  $\rho_{\rm inf}$ and $\rho_{\rm eq}$. But the situation is completely different as regards $\rho_\Lambda$. We have no clue what determines the astonishingly small but non-zero numerical value of the cosmological constant characterized by the number (in natural units with $c=1, \hbar=1$):
  \begin{equation}
  (\rho_{\Lambda} L_P^4)\approx 1.1 \times 10^{-123}
  \end{equation} 
   The late time evolution of the universe is characterized by the cosmological constant $\Lambda$, and the four constants ($\Lambda, G, \hbar, c$) describing nature thus lead to the dimensionless combination 
  \begin{equation}
  \Lambda \left(\frac{G\hbar}{c^3}\right)=8\pi \rho_\Lambda L_P^4\approx 2.8 \times 10^{-122}
  \end{equation}
  which is probably the smallest non-zero number relevant to physics! This issue is well-known and has often been thought of as \textit{the} most fundamental problem in theoretical physics today. 
  
  Incidentally, there is another small number which has not acquired the notoriety it probably deserves. You can easily verify that:
  \begin{equation}
  \rho_{\rm eq}L_P^4 \approx 2.5 \times 10^{-113}
  \end{equation}
  is also extraordinarily small. (This fact comes as a surprise many cosmologists!) How come cosmology literature is totally silent on the fine-tuning problem of $\rho_{\rm eq}$ but gives such a bad press to $\rho_\Lambda$? Do you really believe that $10^{-113}$ is \textit{not} fine-tuning but $10^{-122}$ is? That would be a pretty ridiculous stand to take. When confronted with the smallness of $\rho_{\rm eq}L_P^4$ most people distinguish it conceptually from the smallness of $\rho_\Lambda L_P^4$ by giving two arguments: (a) We \textit{do} have a hope of determining $\rho_{\rm eq}L_P^4$ from high energy physics through \eq{heprhoeq} but we have no clue as to how to go about determining $\rho_\Lambda L_P^4$. (b) It is more likely that $\Lambda$ has ``something to do'' with $L_P^2$ than with $\rho_{\rm eq}$. These arguments may not sound very satisfactory but they do highlight the importance of the cosmological constant problem as something unique.
  
  So, our universe appears to be a hastily put together, make-shift job, using the three densities $\rho_{\rm inf}, \rho_{\rm eq}, \rho_\Lambda$ which \textit{have no relation to each other either conceptually or numerically}. The fact that you can build such an ad-hoc universe is closely related to the theoretical feature I mentioned in Sec.\ref{sec:allok}. The current theoretical framework which uses \textit{only} Einstein's equations is insufficient to constrain  the matter sector. You can build universes with any set of three numbers ($\rho_{\rm inf}, \rho_{\rm eq}, \rho_\Lambda$) and there is no theoretical principle constraining their values. Clearly we need some \textit{additional} theoretical principle to supplement the gravitational field equations if we have to make sense out of these numbers.
  
  Having described the strangeness of the  numerical values of ($\rho_{\rm inf}, \rho_{\rm eq}, \rho_\Lambda$), let me show you something \textit{still stranger.}
    I  invite you to form a specific, dimensionless, number $I$ out of these three densities by the definition:
\begin{equation}
I= \frac{1}{9\pi} \, \ln \left( \frac{4}{27} \frac{\rho_{\rm inf}^{3/2}}{\rho_\Lambda\,\rho_{\rm eq}^{1/2}}\right) 
\label{strange1}
\end{equation} 
and evaluate its numerical value by plugging in the values in \eq{eqn4.13} and \eq{eqn4.14} for $\rho_{\rm eq}$ and  $\rho_\Lambda$ respectively and taking $\rho_{\rm inf}^{1/4} = 10^{15}$ GeV. Surprisingly enough, you will get
\begin{equation}
I   \approx 4\pi \left( 1 \pm \mathcal{O} \left(10^{-3}\right)\right)
\label{strange2}
\end{equation} 
\textit{That is, $I = 4\pi$ to an accuracy of one part in thousand} for the  values of parameters determined from observations and considered reasonable by cosmologists.  This should make you wonder why the right hand side of \eq{strange1} has such a nice value as $4\pi$ since it is not often that  such strange things happen. 

Later in this article (see Sec. \ref{sec:cosmin}),  I will show  that: (i) the right hand side of \eq{strange1} can actually  be interpreted, in a well-defined manner, as the amount of of cosmic information accessible to an eternal observer and (ii) the reason it is $4\pi$ has to do with the quantum microstructure of spacetime.\footnote{It is an \textit{observational fact}  that $I$ defined via \eq{strange1} has a numerical value $4\pi$ for our universe. You need to be a true believer in coincidences if you think such a result can be completely ignored as ``just one of those things''!} 
Obviously, turning this  around and arguing that $I=4\pi$ from theoretical considerations, one can determine the numerical value of \cc\ in terms of the other cosmological parameters, $\rho_{\rm eq}$ and $\rho_{\rm inf}$ which --- eventually --- will be determined from the high energy physics.
But, as I said before, this would require a shift in the theoretical paradigm and \textit{cannot} be done within the conventional approach to cosmology. I will come back to this in Sec. \ref{sec:altpara}. 

\section{Come back aether, all is forgiven?}

Our universe selects a preferred Lorentz frame with respect to which you can measure the absolute velocity of your motion. In a few decades your car will be equipped with a gadget which will couple to the CMB, detect its dipole anisotropy and will tell you your velocity vector with respect to the absolute rest frame of the universe in which the CMB is homogeneous and isotropic. Operationally, to a limited extent, this is no different from the good old aether which was providing an absolute reference frame for defining the state of rest. In fact, the universe also provides you with an absolute \textit{time} coordinate in terms of the CMB temperature. If you specify that you are using a coordinate system in which the CMB is homogeneous and isotropic and the CMB temperature is, say 30 K, you have uniquely specified your Lorentz frame (with the only residual symmetry allowed being that of spatial rotations and spatial translations.)

The field equation of GR, of course, is generally covariant and does not select out any coordinate system --- and, in fact, it is invariant under a much larger group than just the Lorentz group. A \textit{specific solution} to this field equation need not possess the full symmetry of the equation, which is a rather trivial and well known fact. To obtain any specific solution, we need to specify $T^a_b$ which could bring in a natural coordinate system. For example, the metric around Sun has the simplest description if you use a spherically symmetric coordinate system with its origin at the center of the Sun.  Similarly, if $T^a_b$ is spatially homogeneous and isotropic, it is probably simplest to describe the universe using the coordinates in \eq{eqn4.4} or \eq{eqn4.6}. This fact, by itself, is not a cause for surprise or concern.\footnote{More formally, in quantum field theoretic language, we assume that the local vacuum state is Lorentz invariant in the suitable limit. The universe with matter and CMBR is interpreted as a highly excited state of the vacuum which does not have this symmetry.}

Neither does the existence of cosmic `aether' in the form of the CMB violate special relativity in any way. In fact, you can reinvent special relativity using the CMB observations along the following lines: Let A be a geodesic observer who sees the CMB as isotropic (in a local inertial frame in the Friedmann geometry) and let B be another inertial observer, moving with a boosted velocity $\bf v$ with respect to A, who will see the CMB as anisotropic with a dipole anisotropy $T^{-1}(\theta,v)\equiv\beta(\theta,v)= \beta_0[a(v)+b(v)\cos(\theta-\theta_0)]$.  
By varying the direction and magnitude of her velocity, B can determine the functional forms of $a(v),b(v)$ and the value of $\theta_0$ purely from local observations. She will find that $a(v)=\gamma(v), b(v)=\gamma(v)(v/c)$ where $\gamma^{-1}=\sqrt{1-v^2/c^2}$, involving a parameter $c$ which she will recognize is equal to the speed of light.  By comparing the results in \textit{three} inertial frames A,B and C and careful reverse engineering, one can motivate the standard velocity addition formula in SR involving the relative velocity of B and C. This, in turn, will tell you that the speed of light is the same in all inertial frames which are boosted with respect to the freely falling frame in Friedmann universe  (in which the CMB is isotropic). The rest of SR will follow.\footnote{Unfortunately, text books in SR often create the impression that the absence of an absolute frame of rest is important for the validity of SR. Our universe \textit{does} have an absolute frame of rest but its existence does not violate either SR or GR.} All this is  interesting but does not create any problem. 

But the situation does have a deeper level of subtlety, which raises a conceptual issue. 
The real peculiarity is not  the fact that our universe has a preferred Lorentz frame; \textit{it is the fact that we see no trace of this preference at sub-cosmic scale physics.} To appreciate this issue, you have to recall that the smooth  universe is an approximate entity and its description involves averaging the energy momentum tensor of actual clustered matter over sufficiently large scales. In arriving at the standard cosmological model, we \textit{first} average the matter distribution over sufficiently large scales, determine an average $\langle T^a_b \rangle$ and  \textit{then} solve  $G^a_b =\kappa \langle T^a_b\rangle$ to obtain the Friedmann metric $g_{ab}^{\rm FRW}$. It is assumed that if we had solved the exact equations  $G^a_b =\kappa  T^a_b$, found the exact metric $g_{ab}$ and then averaged it over large scales, we would have found that $\langle g_{ab}\rangle = g_{ab}^{\rm FRW}$ to a sufficient level of accuracy.\footnote{Aside: This fact is used  to raise a bogey in cosmology by someone once in a while but this averaging assumption is \textit{always} implicit in the interpretation of GR field equations. In the Einstein's equations $G^a_b = \kappa T^a_b$, you are expected to specify the matter energy momentum tensor at every event in the spacetime and then solve these differential equations. But you have no way of specifying the energy momentum tensor precisely at any given event where matter is present! For example, suppose you want to determine the metric  inside the Sun. Usually you define $T^a_b$ at an event inside the Sun by treating the matter as a fluid; this $T^a_b$ is an approximate $T^a_b$ obtained by averaging a more exact $T^a_b$ over a region large compared to the mean free path. If you probe matter at still smaller scales you will discover averaging at various scales all the way to, say, quarks and gluons. So, strictly speaking, we are \textit{always} solving the equations $G^a_b =\kappa \langle T^a_b\rangle$. But suppose we had found the metric $g_{ab}$ for the exact $T^a_b$ and averaged the metric over scales large compared to the mean free path etc. Will such an average metric match with the solution of $G^a_b =\kappa \langle T^a_b\rangle$ at appropriate length scales? The entire GR works on the assumption that such an averaging \textit{is} valid in spite of the non-linear nature of Einstein's equations. If you do not assume this, you cannot solve Einstein's equations reliably in any region occupied by normal matter.} 

But what coordinate system should we use to solve the exact equation $G^a_b =\kappa  T^a_b$? The validity of standard GR and general covariance at small scales imply that you could have used any coordinate system you like. If you then do the averaging, your final result will indeed be Friedmann \textit{geometry} but expressed in some strange coordinate system.  To recover Friedmann geometry in the standard Friedmann coordinates,  you should work with a sub-class of all possible coordinate systems while solving the exact equations $G^a_b =\kappa  T^a_b$ and doing the averaging. In other words, if you want the exact metric, averaged over large scales, to exhibit the  symmetries of the Friedmann universe explicitly, then you need to restrict the general covariance at small scales.\footnote{You can do this calculation explicitly in the Schwarzschild-deSitter geometry containing a mass $M$ and the cosmological constant $\Lambda$. If you express the metric in static, spherically symmetric coordinates and average the metric over scales large compared to the gravitational radius of $M$, you will indeed recover deSitter \textit{geometry} but in the static coordinates. You have to make a peculiar coordinate transformation of the Schwarzschild-deSitter geometry before averaging if the averaging has to reproduce the deSitter universe in the standard Friedmann coordinates.}

This shows that the existence of an absolute frame of rest at large scales actually selects out a class of coordinate systems with special properties at small scales. But we see no experimental evidence for such a selection  at small scales. To the extent we can determine experimentally, there is no trace of an absolute rest frame or even a preferred class of frames in sub-cosmic level physics. 
The laboratory scale experiments looking for an absolute frame of rest (without using the CMB) have repeatedly drawn a blank, but WMAP or PLANCK has no difficulty in determining it using the CMB, even locally. Roughly speaking, physics at cosmic scales breaks the general covariance (and even Lorentz invariance) operationally by providing us with an absolute standard of rest; but as we move to smaller and smaller scales, we are left with no trace of the cosmic frame of rest in any other phenomena, and the diffeomorphism invariance of equations holds. This enhancement of symmetry\footnote{I mean the \textit{mathematical} symmetry of the \textit{equations} describing the physics; \textit{not} the \textit{physical} symmetry of the matter distribution. Obviously the matter distribution is more symmetric at large scales than at small scales; this is precisely what prevents us from identifying a special reference frame at small scales, but allows us to do it at large scales!} as we proceed to smaller scales is definitely a peculiarity of our universe which cries out for an explanation; once again, we have no clue how to go about it.
 
 In summary, the fact that the \textit{entire} universe is filled by sources (definitely the CMB, even if you ignore clustered matter) which maintains the large scale homogeneity and isotropy is extremely peculiar. It is this substratum which allows us to define a cosmic rest frame, \textit{purely from observations}.  You could certainly describe the Friedmann geometry in any coordinate system you like and the physics will not change; this is assured by the fact that general relativity respects general covariance. But it also  remains a fact that observers can measure --- and indeed they have measured ---  your absolute velocity with respect to a cosmic rest frame. In fact, paradoxical though it might seem, the Friedmann model provides a \textit{generally covariant} procedure for constructing an \textit{absolute} frame of rest!

\section{The arrow of time, expansion and spontaneous classicalization}\label{sec:timearrow}

I will now raise a question which, at the outset, may sound somewhat strange. \textit{Why does the universe expand and, thereby, give us an arrow of time?} To appreciate the significance of this question, recall that \eq{eqn4.7} is invariant under time reversal $t\to -t$. (After all,  Einstein's equations themselves are time reversal invariant.) 
To match the observations, we have to choose a solution with $\dot a>0$ at some fiducial time $t=t_{fid}>0$ (say, at the current epoch), thereby breaking the time-reversal invariance of the system. This, by itself, is not an issue for a laboratory system. We know that a particular solution to the dynamical equations describing the system need not respect all the symmetries of the
equations. But, for the universe, this \textit{is} indeed an issue. 

To see why, let us first discuss the case of $(\rho+3p)>0$ for all $t$.
The \textit{choice} $\dot a>0$, at any instant of time, implies that we are \textit{postulating} that the universe is expanding at that instant. Then \eq{eqn4.7} tells us that the universe will expand at all times in the past and will have a  singularity ($a=0$) at some finite time in the past (which we can take to be $t=0$ without loss of generality). The  structure of \eq{eqn4.7} prevents us from specifying the initial conditions at $t=0$. So, if you insist on specifying the \textit{initial} conditions and integrating the equations forward in time, you are forced to take $\dot a>0$ at some $t=\epsilon>0$, thereby breaking the time reversal symmetry. The universe expands at present `because' we chose it to expand at some instant in the past.
This expansion, in turn, gives us an arrow of time
with either $t$ or $a$ can be used as a time coordinate.
But why do we have to choose the solution with $\dot a>0$ at some instant?. This is the essence of the so called \textit{expansion problem} \cite{peacock}.
An alternative way of posing the same question is the following: How come a cosmological arrow of time emerges from equations of motion which are time-reversal invariant?

In a laboratory, we can usually take another copy of the system we are studying and explore it with
a time-reversed choice of initial conditions, because the time can be specified by degrees of freedom \textit{external} to the  system. We cannot do it for the universe because we do not have extra copies of it handy and --- equally importantly --- there is nothing external to it to specify the time. So the problem, as described, is specific to cosmology.

So far we assumed that $(\rho+3p)>0$, thereby leading to a singularity. Since  meaningful theories must be nonsingular, we certainly expect a future theory of gravity --- possibly a model for quantum gravity --- to eliminate the singularity [effectively leading to $(\rho + 3p) <0$]. Can such a theory  solve the problem of the arrow of time? This seems unlikely.  To see this, let us ask what kind of dynamics we would expect in such a `final' theory. The classical dynamics will certainly get modified at the Planck epoch, but, away from it, we expect some effective equations (possibly with quantum corrections) to govern the evolution of an (effective) expansion factor. The solutions could, for example, have a contracting phase (followed by a bounce) or could start from a Planck-size universe at $t=-\infty$, just to give two non-singular possibilities. While we do not know these equations or their solutions, we can be confident that they will still be time-reversal invariant because quantum theory, as we know it, is time-reversal invariant. So except through a choice for initial conditions (now possibly at $t=-\infty$), we still cannot explain how the cosmological arrow of time emerges. Since quantum gravity is unlikely to produce an arrow of time, it is a worthwhile pursuit to try and understand this problem in the (semi)classical context.\footnote{A more complicated ``solution" to the arrow of time issue, which is sometimes suggested, is as follows: 
Consider a very inhomogeneous initial condition in some $t=t_i$ hypersurface and let us assume that
certain regions behave like `local' Friedmann models with $\dot a>0$ and other regions have
$\dot a<0$ so that no global arrow of time can be defined from the expansion. Next, assume that the dynamics of these patches are independent of each other and we just happen to be in a patch with 
$\dot a>0$, thereby `solving' the problem. This scenario has  several difficulties. For a generic initial condition, the patches will \textit{not} evolve independently in a nonlinear theory of gravity. Even defining a  `local expansion factor' for a `local patch' without assuming special symmetries is impossible. Such scenarios are often invoked in the context of inflationary models but they do not have  rigorous mathematical justification.}

Given all these, it will be nice  if we can find a simpler way by which the  equations of motion that are time-reversal invariant can lead to  an evolution which singles out an arrow of time. At first sight one might think  this is impossible but one can manage to do it with unbounded Hamiltonians. I will describe the idea with a simple example. Consider an `inverted'  oscillator $q(t)$ obeying the equations of motion
\begin{equation}
\ddot q=\omega^2 q
\label{invosci}
\end{equation} 
This equation is clearly invariant under $t\to -t$ so one would have thought that no arrow of time will emerge from the dynamics, unless we impose it in the initial conditions. The general solution \eq{invosci} is
\begin{equation}
q(t)=q(0)\cosh\omega t+\dot q(0)\omega^{-1}\sinh\omega t
\label{mirone}
\end{equation}
For a generic initial condition, there is no relationship between $q(0)$ and $\dot q(0)$. So at late times (i.e,
$t\gg \omega^{-1}$), we find that one branch of the solution is selected out:
\begin{equation}
q(t)\approx\frac{1}{2}(q(0)+\dot q(0)\omega^{-1})e^{\omega t}\propto e^{\omega t}
\end{equation} 
leading to an ``expansion" and an arrow of time! The solution in \eq{mirone} 
is time-reversal invariant in the sense that $q(t)=q(-t)$ if we let $\dot q(0)\to -\dot q(0)$ when we do $t\to-t$. But once we have chosen a generic solution with some uncorrelated $q(0)$ and $\dot q(0)$, the \textit{late time dynamics} picks out an arrow of time correlating the increase of $q^2(t)$ with the increase of $t$. (Of course, there are special initial conditions like e.g., $q(0)=-\dot q(0)\omega^{-1}$ or $q(0)=0=\dot q(0)$ for which this will not happen, but these are special choices and not generic.)

One can easily show that 
 this behaviour arises for a wide class of 
 Hamiltonians that are unbounded. It is not necessary that the potential energy is unbounded. If the kinetic energy term has the `wrong' sign, so that the Lagrangian has a form like $L=-(1/2)\dot q^2-V(q)$ with a $V$ which is positive and unbounded from above, say, we will again end up with  an instability and the late time evolution of $q$ will give an arrow of time.

Interestingly enough, it was known for decades \cite{bryce} that the expansion factor $a(t)$ does have such a wrong sign in the kinetic energy term in the Hilbert action and hence represents an unstable mode. The above interpretation  suggests that \textit{it is this cosmic instability which we call expansion,} and for timescales larger than the Planck time, it picks out an arrow of time. Clearly, the same feature will occur even in any effective theory describing the (semi)classical gravity once $a(t)$ acquires an unstable dynamics. This is guaranteed to happen because any sensible quantum cosmological model will approach the Friedmann model at times larger than the Planck time and --- in the Friedmann limit --- $a(t)$ is an unstable model.

This instability  has relevance for another peculiarity related to the early, quantum regime of the universe which does not seem to have attracted the attention it deserves. It is generally believed that the very early phase of the universe needs to be described quantum mechanically using the principles of quantum gravity, because during these earliest moments, the length scale associated with the curvature will be comparable to the Planck scale. This, however, raises a conundrum not encountered  elsewhere in physics.  \textit{How come the universe, which started out as a quantum mechanical system,  became classical spontaneously   as it evolved?!} 

To see this issue in proper context, you need to introduce some theoretical structure which can tell you the `level of classicality' of a system. Many such definitions can be given, each suited for specific systems  (for a sample of ideas and  references to previous work, see \cite{spcla,clas1,clas2}). Most of them use the idea that a classical system follows a sharply peaked trajectory in \textit{phase space} of the form $p= f(q)$ while a quantum system will not exhibit correlations between $q$ and $p$. You can choose any sensible descriptor of classicality and ask whether a quantum mechanical system can become classical spontaneously. Just for illustration, consider the Wigner function $W(q, p, t)$ built from  a wave function $\psi(t, q)$ describing a quantum system evolving under the action of  a Hamiltonian $H(q,p)$. We would like to know what kind of Hamiltonians  will ensure that, as $t\to \infty$, the Wigner function gets sharply peaked on a classical trajectory in the phase space. 

The answer is  surprising: If the Hamiltonian is bounded, then the system cannot evolve spontaneously to classical behaviour. In other words, if you want to start with a quantum  universe and ensure that the evolution takes it to a classical universe, the effective Hamiltonian describing such an evolution cannot be bounded; rather, it should exhibit an instability from a conventional point of view. As I said before, many of the toy Hamiltonians describing the reduced phase space of gravity do have one degree of freedom --- precisely the one corresponding to the expansion factor $a(t)$ --- which comes up with a wrong sign for the kinetic energy. The above result tells you that the wrong sign is indeed the right sign if the universe has to become classical on its own.

This result is very special to cosmology. Hamiltonians with negative kinetic energy terms are taboo in laboratory scale physics for good reasons; you do not want run-away situations in the lab. But one can easily accommodate such an instability in the behavior of the universe. \textit{The cosmic expansion itself is just a run-away solution fed by the instability.}
This fact also ties up two concepts: (i)  the origin of the cosmic arrow of time and (ii) the spontaneous quantum to classical transition made by the universe during its evolution. Normal  systems in the lab do not spontaneously evolve into a more and more classical state  as time evolves. But this is precisely what systems with unbounded Hamiltonians do. For example, the inverted harmonic oscillator in \eq{invosci}, treated as a quantum system, does become more and more classical as it evolves.

A more general description of this results is as follows:  Decompose the spatial 3-metric in the form $g_{\alpha\beta}=a^2h_{\alpha\beta}$ with $\mathrm{det}\ h=1$ as a gauge condition. Then, using the form of the Einstein-Hilbert Lagrangian, one can show that while the kinetic energy term for $a(t)$ has the wrong sign, the other degrees of freedom, represented by $h_{\alpha\beta}$, have the correct sign. In other words, it is only the overall scale factor of the 3-metric which has an instability. Hence it is this degree of freedom which turns classical first during the evolution. If these features are preserved in the effective quantum-corrected description of gravity, then we can hope to have an explanation for a broader question: Why is the classical universe described by a single dynamical degree of freedom $a(t)$, rather than by, say a Bianchi type-I model with three degrees of freedom?

 Unfortunately, the results obtained so far with regards to this question are limited to simple toy models. There is no assurance that they will hold in a more general context of quantum gravity. But if they do, it tells you that the quantum gravitational description of spacetime should contain the seeds for an instability which is rather unexpected. The standard models for high energy physics and many candidate models of quantum gravity shy away from working with unbounded Hamiltonians because they are mathematical nightmares. What is more, this instability doesn't seem to do much at smaller scales in the universe: We do not see spontaneous nucleation of mini-universes all over space today. It is difficult to embed a quantum cosmological model within the broader context of quantum gravity such that everything will work out fine for the large scale universe but physics at smaller scales will not have any unwanted instabilities. Once again, it appears that the universe is a rather special system and not just a specific solution to a certain more general gravitational theory.

\section{An alternative paradigm for cosmology}\label{sec:altpara}

Let me summarize the discussion so far before proceeding further: The conventional approach to cosmology, which describes the universe using a specific solution to the gravitational field equations, leads to the following issues:

\begin{enumerate}

 \item Einstein's equation, in general, does not put any constraint on $T^a_b$ except requiring $\nabla_a T^a_b =0$. In the context of cosmology, this allows you to accommodate any evolutionary history for the universe with a suitable choice for the equation $p = p(\rho)$. The fact that we will  be able to probe the matter sector only up to a finite energy scale in the lab at any given time, while the energy scale close to the big bang can be arbitrarily high (if $p>0,\rho>0$),  makes the early evolution of the universe under-determined both in principle and in practice.

 \item Observations suggest that the evolution of our universe is well approximated by the differential equation \eq{eqn4.11} containing three constant parameters $\rho_{\rm inf}, \rho_{\rm eq}, \rho_\Lambda$ with two --- apparently unconnected --- epochs of accelerated expansion. These three parameters, which constitute the signature of our universe, do not seem to have any conceptual or numerical relationship. In other words, our universe is built using three unrelated, ad-hoc numbers.
 
 \item It is possible to construct a rather strange combination of these three densities and define a quantity $I$ (see \eq{strange1}) which has the numerical value $4\pi$ to the precision of 1 part in 1000.  This (four) ``pi in the sky'' demands an explanation, which is difficult to conceive of within the context of the conventional approach because the matter sector is completely unconstrained. (``Everything is allowed in cosmology.'')

 \item The cosmos, at very large scales, provides us with an absolute frame of rest (in which the CMB is isotropic) and an absolute time coordinate (in the form of the temperature of the CMB). You can measure your absolute motion using the dipole anisotropy of the CMB, which acts like a cosmic aether. However, we see no experimental trace of the existence of such an absolute standard of rest at sub-cosmic scales (if we do not use the CMB). In other words, sub-cosmic scale physics appears to be invariant under a much larger group (viz., the general coordinate transformation group) than the very large scales which define a cosmic aether and a preferred coordinate system. This enhancement of symmetry at small scales vis-a-vis the largest scales is intriguing.

 \item Cosmic evolution introduces an arrow of time even though Einstein's equations, like the rest of physics, are invariant under time reversal. (This arrow of time arises through dynamics rather than through statistical coarse graining etc.) While the unstable mode, corresponding to the expansion factor $a(t)$, might have something to do with this, the details are still uncertain. In particular, this explanation requires the effective Hamiltonian describing the cosmos to be unbounded; it is not clear how such a feature could be embedded in a quantum gravitational model without affecting small scale physics. Once again, we see a conceptual tension between the description of sub-cosmic scale physics and the dynamics of the large scale universe. 
 
 \item The universe is probably the only system known to us which made a spontaneous transition from  quantum mechanical behaviour  to  classical behaviour --- i.e, treated as a dynamical system, it became more and more classical as time went on. Systems with bounded Hamiltonians cannot do this. This suggests a relation between spontaneous transition to classicality and the arrow of time, possibly again through the ``wrong sign'' for the kinetic energy associated with the expansion factor. While this could very well be the reason, the details remain to be worked out. 

 \end{enumerate}

I believe the conundrums described above provide sufficient motivation to look for an alternative paradigm to describe the cosmos. In this last part, I will outline such a paradigm and how it addresses at least one of the crucial issues, viz., the problem of the cosmological constant. 

The conventional approach  begins by assuming the validity of GR to describe the evolution of spacetime and then obtain a specific solution to the field equation to describe the evolution of the large scale universe. There is, however, considerable evidence to suggest that the field equations of gravity themselves  have only the  conceptual status similar to  the equations describing an elastic solid or a fluid (see e.g., \cite{grtp,tpreviews}). In the alternative perspective which emphasizes this feature, gravity is the thermodynamical limit of the statistical mechanics of the underlying spacetime degrees of freedom (the `atoms of space'). The field equations are obtained from a thermodynamic variational principle which is similar to extremizing a thermodynamic potential to obtain the equilibrium state of the normal matter. The validity of the GR field equations then correspond to a maximum entropy configuration of the atoms of space. 

The evolution of spacetime itself can be described in a \textit{purely thermodynamic language} in terms of suitably defined degrees of freedom in the bulk and boundary of a 3-volume. It can be shown \cite{grtp}  that 
 the evolution of geometry, interpreted as the heating and cooling of null surfaces, is described by the equation:
\begin{equation}
\int_\mathcal{V} \frac{d^3x}{8\pi L_P^2}\sqrt{h} u_a g^{ij} \pounds_\xi p^a_{ij} = \epsilon\frac{1}{2} k_B T_{\rm avg} ( N_{\rm sur} - N_{\rm bulk})
\label{evlnsnb}
\end{equation} 
where
\begin{equation}
 N_{\rm sur}\equiv\int_{\partial \mathcal{V}} \frac{\sqrt{\sigma}\, d^2 x}{L_P^2};\quad
N_{\rm bulk}\equiv \frac{|E|}{(1/2)k_BT_{\rm avg}}
\end{equation} 
 are the degrees of freedom in the surface $\partial \mathcal{V}$ and bulk $\mathcal{V}$ of a 3-dimensional region  and $T_{\rm avg}$ is the average Davies-Unruh temperature \cite{du1,du2}  of the boundary. 
The  $h_{ab}$ is the induced metric on the $t=$ constant surface, $p^{a}_{bc} \equiv -\Gamma^{a}_{bc}+\frac{1}{2}(\Gamma^d_{bd}\delta^{a}_{c}+\Gamma^d_{cd}\delta^{a}_{b})$, and $\xi^a=Nu^a$ is the proper-time evolution vector corresponding to observers moving with four-velocity $u_a  = - N \nabla_a t$. The factor $\epsilon=\pm1$ ensures the correct result for either sign of the Komar energy $E$.
The time evolution of the metric in a region (described by the left hand side of \eq{evlnsnb}),  can be interpreted \cite{A21} as the heating/cooling of the spacetime and arises because $N_{\rm sur}\neq N_{\rm bulk}$. In any \textit{static} spacetime \cite{tpsurf}, on the other hand, $\pounds_\xi(...)=0$, leading to ``holographic equipartition'': 
$N_{\rm sur}= N_{\rm bulk}$.  

Equation (\ref{evlnsnb}) translates the gravitational dynamics  into  the  thermal evolution of the spacetime.
 The validity of \eq{evlnsnb} for all observers (i.e., foliations) ensures the validity of Einstein's equations.
 I stress that, even though \eq{evlnsnb} describes a time evolution, it is obtained from an extremum condition for a thermodynamic variational principle and represents the thermodynamic equilibrium between matter degrees of freedom and \md. 
 
 In the specific context of cosmology, one can write a similar but simpler equation of the form \cite{tpemeuniv}:
 \begin{equation}
 \frac{dV_H}{dt}=L_P^2(N_{sur}-N_{bulk})
 \label{evl1}
\end{equation}
where $V_H= (4\pi/3)H^{-3}$ is the volume of the Hubble sphere, $N_{\rm sur}= A_H/(L_P^2)=4\pi H^{-2}/L_P^2$ is the number of \md\ on the Hubble sphere, $N_{\rm bulk}=-E/[(1/2)k_BT]$ is the equipartition value for the bulk degrees of freedom corresponding to the Komar energy $E$ contained in the Hubble sphere, and $T=(H/2\pi)$ is the Hubble temperature. (I have assumed $E<0$ to describe the current accelerated phase of the universe; otherwise one needs to flip suitable signs to keep $N_{bulk}>0.$) This equation is equivalent to the second equation in \eq{eqn4.7}. In fact one can also rewrite the first equation in \eq{eqn4.7} in thermodynamic language as an energy balance relation:
\begin{equation}
 \rho V_H=TS 
 \label{evl2}
\end{equation}
where $S=A_H/(4L_P^2) = \pi H^{-2}/L_P^2$ is the entropy associated with area of the Hubble sphere making $TS$ in the right hand side the heat energy of the boundary surface. This equation tells you that the total energy within the Hubble sphere is equal to the heat energy of the boundary surface.

But if this is indeed the correct description of the \md, then the field equations of GR, representing some kind of thermodynamic equilibrium between matter and the \md, cannot be universally valid.  
 Recall that, in the case of standard fluid mechanics, we may have to abandon the thermodynamic description in two different contexts. First, if we probe the fluid at scales comparable to the mean free path, you need to take into account the discreteness of molecules etc., and the fluid description breaks down. Second, a fluid simply might not have reached local thermodynamic equilibrium at the scales (which can be large compared to the mean free path) we are interested in. In the first case, the `fluid' description itself breaks down; in the second case we do have a continuum description of the fluid, but it needs to be studied using non-equilibrium kinetic theory. 

Something analogous happens in the description of gravity. 
The \md\ could have reached the maximum entropy configuration at sub-cosmic scales, making the standard field equations of gravity valid at these scales (say at scales $10^6 L_P\lesssim x\lesssim H^{-1}$). Equation \ref{evlnsnb} (which is identical to $G^a_b=\kappa T^a_b$ but expressed in a thermodynamic language) holds at these scales. For scales close to $L_P$, the discrete nature of spacetime has to be taken into account and this is similar to probing a fluid at scales comparable to the mean free path; we do not yet know how to do this, which is the usual problem of quantum gravity. But it is also possible that the \md\ have not reached the maximum entropy configuration at very large scales comparable to the \textit{horizon} scale (which is much larger than the scale of the \textit{Hubble radius} in the RD and MD phases). At these scales we again expect \eq{evl1} and \eq{evl2} to be modified because the \md\ are not in the maximum entropy configuration. This is similar to the situation in non-equilibrium thermodynamics for normal fluids. In such a description, the symmetry of Einstein's equations, viz. general covariance emerges when the \md\ reach the maximum entropy configuration at the intermediate scales. At very large scales, this `equilibrium' has not yet been achieved and 
 the universe, at very large scales,  picks out a cosmic frame of rest. (Of course,  we also do not know how to introduce the concept of general covariance in a meaningful way close to Planck scales; but that is a different --- and more well-known --- story.)\footnote{A very figurative analogy is provided by a large chunk of ice containing a point source of heat inside. The heat melts ice around it creating a  region containing  water, which 
 expands, maintaining thermodynamic equilibrium. The degrees of freedom in the form of water have a higher degree of symmetry (rotational invariance), compared to the degrees of freedom locked up in the ice-lattice. In the case of the universe, the expansion  leads to the emergence of space \cite{tpemeuniv} with the \md\ reaching the maximum entropy configuration. This region exhibits a higher degree of symmetry (the general covariance of \eq{evlnsnb}, which as I said, is just $G^a_b=\kappa T^a_b$), compared to the larger scales of the universe.}

So what could be the additional ingredient we need to introduce into the standard GR? \textit{I believe \cite{tpreviews} this has to do with the concept of information stored in the spacetime and its accessibility  by different observers.}

A key feature of gravity is its ability to control the amount of information accessible to any given observer.  A well-known example of this idea arises in the physics of black holes. 
More generally, the lack of access to spacetime regions leads to a configurational entropy related to the \md. Over decades, we have come to realize \cite{info} that information is a physical entity and that anything which affects the flow and accessibility of information will have direct physical significance.
 One consequence of such a paradigm is that matter and geometry will be more closely tied together (through the information content) than in the conventional approach. 
You should not be able to build an ad-hoc universe with randomly chosen values for the three densities $\rho_{\rm inf}, \rho_{\rm eq}, \rho_\Lambda$. \textit{We would expect, for example,  a  relationship connecting the late time accelerated expansion with the early inflationary phase through the information content of spacetime.}

It turns out that, by applying this   idea  to the cosmos in a specific manner, you can solve \textit{the} deepest mystery about our universe, viz., the small numerical value ($\Lambda L_P^2 \approx 10^{-122}$) of the \cc, $\Lambda$. 
The $I$ in \eq{strange1} measures (in a way defined more precisely below) the amount of information accessible to an eternal observer in our universe. If $\rho_\Lambda =0$, such an observer can observe all of spacetime and can acquire an infinite amount of information. But when $\rho_\Lambda \neq 0$, the information accessible to the observer is finite and is related to $\rho_\Lambda$ through \eq{strange1}. So, if you have an independent way of fixing $I$, then you can use \eq{strange1} to express $\rho_\Lambda$ in terms of the other two densities. The value of $I$ is fixed by the following fact: It turns out that, \textit{the spacetime becomes effectively 2-dimensional close to Planck scales},  irrespective of the dimension exhibited by the spacetime at large scales \cite{deq2}. This, in turn, implies that the basic unit of information stored in the \md\ is given by $A/L_P^2= 4\pi$ where $A=4\pi L_P^2$ is the area of the 2-sphere with radius $L_P$. This unit of quantum gravitational information is what appears in \eq{strange2} and allows us to determine the numerical value of the cosmological constant. Let me now fill in some details of this result.\footnote{These results are based on unpublished work done in collaboration with H. Padmanabhan.}

\section{Cosmic Information and the cosmological constant}\label{sec:cosmin}

Let us begin by recalling how the existence of a non-zero \cc\ prevents an eternal observer $O$ (i.e., an observer whose world line extends to $t\to \infty$ and who makes observations at very late times) from acquiring information from the far reaches of our universe. Let $x(a_2,a_1)$ be the comoving distance traveled by  light  between the epochs $a=a_1$ and $a=a_2$ with $a_2>a_1$ in the standard Friedmann model with expansion factor $a(t)$. This is given by:
\begin{equation}
 x(a_2,a_1) = \int_{t_1}^{t_2} \frac{dt}{a(t)} = \int_{a_1}^{a_2} \frac{da}{a^2H(a)}
\end{equation} 
 So the  comoving [$x_\infty(a)$] and proper [$r_\infty (a)$] sizes of the regions of the universe at an epoch $a$, from which $O$ can receive signals at very late times, are given by \cite{fb}:
\begin{equation}
x(\infty,a)\equiv x_\infty(a) = \int_a^\infty \frac{d\bar a}{\bar{a}^2 H(\bar a)}; \qquad r_\infty (a)=ax_\infty(a) \equiv a \int_a^\infty \frac{d\bar a}{\bar{a}^2 H(\bar a)}
\label{one}
\end{equation} 
The behaviour of $x_\infty(a)$ and $r_\infty (a)$ depend crucially on whether the \cc\ is zero or non-zero.
If $\Lambda =0$ and the universe is dominated by, say,  matter at late times, then  $H(a) \propto a^{-n}$, with $n>1$ at late times. Then, both these integrals will diverge at the \textit{upper} limit as $t\to \infty$, irrespective of the behaviour of the universe at earlier epochs. 
So, in a universe with $\Lambda =0$, information from the infinite expanse of space will be accessible to the eternal observer at late times; there is no blocking of information.

 If $\Lambda \neq 0$ and $H(a) \to H_\Lambda = $ constant at late times, then the situation is  different. In that case, both the integrals in \eq{one}
are finite at the upper limit and an eternal observer have only access information from a \textit{finite} region of space at an epoch $a$, irrespective of how long she waits.  The amount of accessible Cosmic Information (which I call ``CosmIn'') is now reduced from an infinite amount to a finite value, say $I_c$, as a \textit{direct consequence} of the fact that $\Lambda \neq 0$. \textit{It is, therefore, reasonable to expect that the  numerical value of $\Lambda$ should be related to $I_c$ with $I_c$ decreasing with increasing  $\Lambda$.} I will now derive this relation. 

 Consider a universe (like ours) with three distinct phases of evolution: (i) At very early times, the universe was in a state of inflation with $H(a) = H_{\rm inf} = $ constant. (ii) At some point $a = a_{\rm rh}$, the inflation ends; the universe reheats and becomes radiation-dominated. The radiation dominated phase goes on till $a=a_{\rm eq}$ which is the epoch of radiation-matter equality. For  $a_{\rm eq} \lesssim a \lesssim a_\Lambda$, the universe is matter-dominated. (iii) Later on, at some point,  $a \gtrsim a_\Lambda$,  the \cc\ starts to dominate the expansion of the universe. We will rescale the expansion factor such that $a_{\rm eq} =1$, and use \eq{eqn4.11} to describe the evolution of the universe. 
 I will also assume instant reheating at $a= a_{\rm rh}$ for simplicity.\footnote{For our universe, observations tell us that $a_{\rm rh}\approx 7.4\times 10^{-25}, a_\Lambda\approx2.8\times 10^3,$ if we set $a_{\rm eq}=1$.}

\begin{figure}
\begin{center}
\scalebox{0.48}{\input{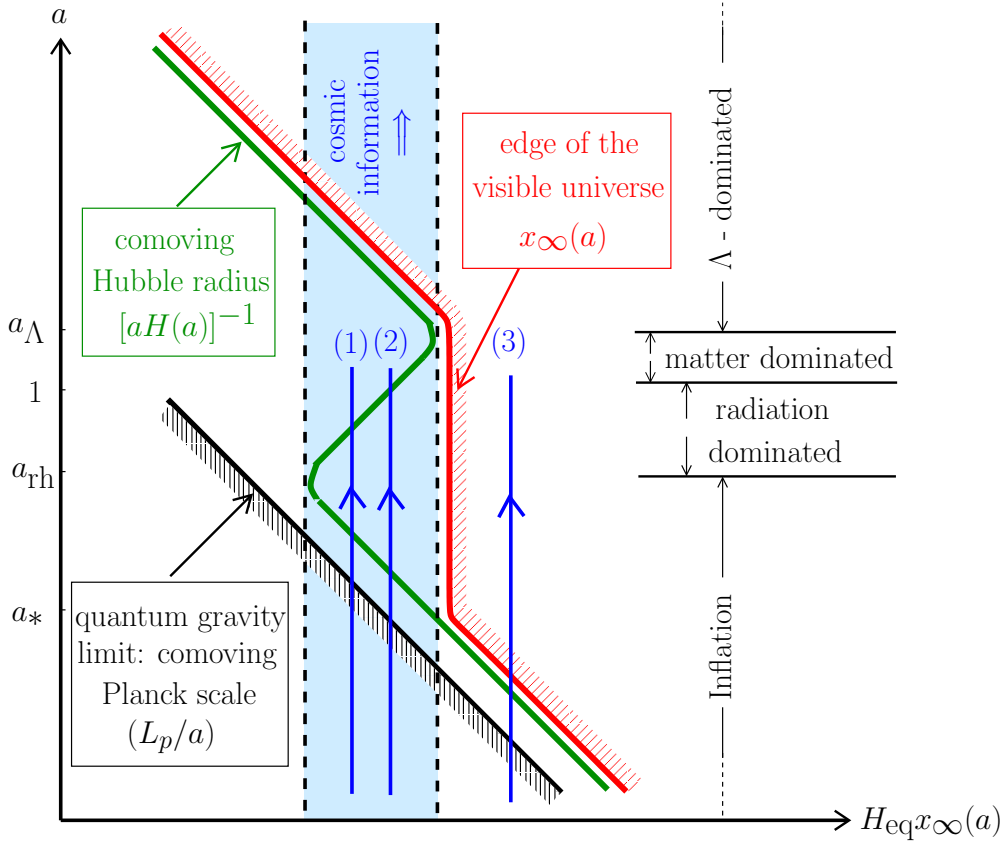}}
 \caption{The different length scales in a universe with an inflationary phase and a non-zero \cc. The red curve is the maximum comoving size of a region from which signals can reach an observer at very late times. The information in the  shaded region to the right of the red curve is not accessible to an observer even if she waits till eternity. 
The green curve is the comoving Hubble radius. The slanted black curve is the comoving scale corresponding to the Planck length and the shaded region below this black line is dominated by quantum gravitational effects. The vertical lines are different proper length scales which cross the Hubble radius and the horizon. The two lines, marked 1 and 2, leave the Hubble radius during inflation and re-enter it during the radiation/matter dominated epoch. These remain within the horizon of the observer at the origin (red curve) and are visible to her at, say, $a=a_{\rm rh}$. The line marked 3 corresponds to  a proper length scale which goes out of, not only the Hubble radius, \textit{but also the horizon} and thus become inaccessible to the observer at, say, $a= a_{\rm rh}$. So the relevant part of  the cosmic information is contained within the blue vertical band, between the two vertical lines which are tangential to the comoving Hubble radius at its turning points. The arrows at the top of the band denote the direction of flow of the cosmic information.}
\label{fig:lengthscales}
\end{center}
\end{figure}

The  geometrical features related to $x_\infty(a)$ and other relevant length scales are depicted in Fig. \ref{fig:lengthscales}.
  The green curve is the \textit{comoving} Hubble radius $d_H(a)/a \equiv 1/ aH(a)$. It  decreases (as $1/a$) during the inflationary phase, reaching a minimum at $a= a_{\rm rh}$; it then increases as $a^2$ in the radiation-dominated phase and as $a^{3/2}$ in the matter-dominated phase, attaining a maximum around $a \approx a_\Lambda$;  in the $\Lambda$-dominated phase, it again decreases as $1/a$.
The red curve  denotes $x_\infty(a)$  obtained by evaluating the integral in \eq{one} and represents the visibility limit.
During the  $\Lambda$-dominated phase, this curve closely tracks the comoving Hubble radius ($x\approx a_\Lambda^{3/2}/a$) but soon
becomes  vertical to a high degree of approximation. During the matter and radiation dominated phases (i.e, during $a_\Lambda \gtrsim a \gtrsim a_{\rm rh}$) the $x_\infty(a)$ is approximately constant --- varying  just by a factor 3 (from $\sim a_\Lambda^{1/2}$ at $a=a_\Lambda$ to $\sim 3 a_\Lambda^{1/2}$ at $a=a_{\rm rh}$) when $a$ varies by a factor $\sim 10^{28}$.  During the inflationary phase, $x_\infty(a)$ again tracks $d_H/a$ asymptotically, 
with an approximate behaviour $x_\infty(a) \approx [3 a_\Lambda^{1/2} - a_{\rm rh}] + a_{\rm rh}^2/a$. 
As I mentioned, the region of space from which an eternal observer can acquire information is finite for all finite $a$ if the \cc\ is non-zero.

 Our next task is to quantify the amount of  cosmic information that is actually accessible to the eternal observer. To do this, recall that a comoving scale  $x= $ constant is associated with a proper length scale $r = a (t) x$. The proper length scales (e.g.,  the wavelengths of  modes)
will get stretched exponentially during the inflation, and will exit the Hubble radius. (One can think of $\lambda(t)=a(t)\lambda_0$ as a physical length being stretched by expansion. You can equally well think of it as the proper length associated with a geodesic labeled by a comoving scale $x_0=\lambda_0$. I will use the former terminology since it is more familiar to cosmologists.) After remaining outside the Hubble radius for some time, some of them will re-enter the Hubble radius during the matter/radiation dominated  epoch. (Two such modes are shown as (1) and (2) in Fig. \ref{fig:lengthscales}.)
In contrast, the  mode marked as (3) will leave the Hubble radius but will \textit{never} re-enter it. Such modes  cross not only the Hubble radius but also the horizon (red line) and become invisible to the eternal observer at, say, the epoch of reheating $a=a_{\rm rh}$. So the modes  relevant to us are confined to  those between the two dotted horizontal lines which are tangential to the Hubble radius at its turning points. The total number of such modes (or, equivalently, geodesics) is  a  measure of the information content $I_c$. 

Let us compute how many   modes cross the Hubble radius during the inflationary phase between $a=a_*$ and $a=a_{\rm rh}$.  
Since the deSitter space is invariant under time translation, the \textit{rate} at which the modes leave the Hubble radius will be a constant. So the number of modes $I(a_2,a_1)$ which cross the Hubble radius during an interval $a_1<a<a_2$ will be proportional to $H(t_2-t_1)$. So the total number of modes which cross the Hubble radius during the inflationary epoch will be  proportional to  $N_e\equiv H \Delta t $, where $\Delta t$ is the relevant duration in the inflationary phase. (Here
$N_e$ denotes the number of e-foldings in the interval $\Delta t $.) Therefore,  the CosmIn  is given by:
\begin{equation}
 I_c  \propto N_e
\label{ic1}
\end{equation}
and all we need to do is to fix the proportionality constant using a suitable measure. This measure can be introduced as  
follows:
The number of modes $dN$ within the comoving Hubble volume $V_H(a) = (4\pi/3) (aH)^{-3}$ with wave numbers in the range  $d^3k$ is given by $dN = V_H(a) d^3k/(2\pi)^3=V_H(a)dV_k/(2\pi)^3$ where $dV_k=4\pi k^2 dk$. 
A mode with the comoving wave number $k$ will leave the Hubble radius when $k=k(a)\equiv a H (a)$. So the modes with wave numbers within the range  ($k,k+dk$), where $dk= [d(aH)/da]\, da$, will exit the Hubble radius 
in an interval ($a, a+da$). Therefore, the number of modes that cross the Hubble radius during the interval $a_1<a<a_2$ is given by 
\begin{equation}
N(a_2,a_1) = \int_{a_1}^{a_2} \frac{V_H(a)}{(2\pi)^3} \, \frac{dV_k[k(a)]}{da} \, da = \frac{2}{3\pi}\ln \left( \frac{a_2 H_2}{a_1H_1}\right)
\end{equation} 
(This result, of course,  is applicable for any  $a(t)$; not just in the inflationary phase.) During inflation, when $a(t) \propto \exp(H_{\rm inf} t)$, this expression reduces to $(2/3\pi) \ln(a_2/a_1)$ which tells us that the proportionality constant in \eq{ic1} is $(2/3\pi)$. We thus find the value of CosmIn to be:
\begin{equation}
 I_c = \frac{2}{3\pi} N_e = \frac{2}{3\pi} \ln \left(\frac{a_{\rm rh}}{a_*}\right)
\label{ic2}
\end{equation} 
We can relate the ratio $a_{\rm rh}/a_*$
to the three densities $\rho_\Lambda, \rho_{\rm eq}$ and $\rho_{\rm inf}$ which will give $a_{\rm rh}/a_* \propto (\rho_{\rm inf}/\rho_{\rm eq})^{1/4} \, (\rho_{\rm eq}/\rho_\Lambda)^{1/6}$. To determine the proportionality constant, we need to evaluate the turning point of the $d_H(a)/a$ curve near $a = a_\Lambda$ which requires solving a cubic equation. Doing this (for details, see \cite{hptp}), we find that the proportionality constant has the value $(4/27)^{1/6}=2^{1/3}/3^{1/2}$. Substituting into \eq{ic2}, we  achieve our first goal, viz. relating the \textit{non-zero} value of the \cc\ to the \textit{finite} amount of cosmic information accessible to an eternal observer ($I_c$):
\begin{equation}
\rho_\Lambda = \frac{4}{27} \ \frac{\rho_{\rm inf}^{3/2}}{\rho_{\rm eq}^{1/2}} \ \exp \left( - 9 \pi I_c\right)
\label{four}
\end{equation} 
As  expected, the \cc\ vanishes when the information content is infinite ($I_c \to \infty$) and vice-versa.

Equation~(\ref{four}) will determine $\rho_\Lambda$ in terms of $\rho_{\rm inf}$ and $\rho_{\rm eq}$  provided we know the value of CosmIn from some other physical consideration. (I share the hope that   $\rho_{\rm inf}$ and $\rho_{\rm eq}$ will be eventually determined from high energy physics in terms of the  inflationary model and the dark matter content of the universe.) 
To do this, notice that the  modes which exit the Hubble radius  during the inflationary epoch correspond to sub-Planckian scales in the early part of inflation. In Fig. \ref{fig:lengthscales}, the black line indicates the \textit{comoving} length scale $L_P/a$ corresponding to the Planck length $L_P$. The region below this line corresponds to \textit{proper} length scales smaller than the Planck length, and will be dominated by quantum gravitational effects.  The modes  containing the cosmic information cross the comoving Planck length during the earlier stages of evolution and hence carry the imprint of quantum gravitational effects. So we expect $I_c$  to be determined by  quantum gravitational considerations.

It can be shown that the spacetime becomes effectively two-dimensional near Planck scales \cite{deq2} and hence the  unit $I_{\rm QG}$ of quantum gravitational information content of spacetime  is given by the degrees of freedom contained in a 2-sphere of radius $L_P$, viz., 
$I_{\rm QG} = 4\pi L_P^2 / L_P^2 = 4\pi$.
Therefore the natural numerical value for $I_c$ can be taken to be:
\begin{equation}
 I_c = I_{\rm QG} = 4\pi
\end{equation} 
Substituting  into \eq{four}, we get a remarkable formula for the \cc\
\begin{equation}
\rho_\Lambda = \frac{4}{27} \ \frac{\rho_{\rm inf}^{3/2}}{\rho_{\rm eq}^{1/2}} \ \exp \left( -36\, \pi^2\right)
\label{five}
\end{equation} 
If we use the typical values $\rho_{\rm inf} = (1.2 \times 10^{15}$ GeV)$^4, \rho_{\rm eq} = (0.86$ eV)$^4$, we get $\rho_\Lambda = (2.2 \times 10^{-3}$ eV)$^4$ which agrees well with the observed value! 
That is, the idea that the cosmic information content accessible to an eternal observer, $I_c$, is equal to the basic quantum gravitational unit of information $I_{\rm QG} = 4\pi$, determines the numerical value of the \cc\ correctly.
Let me conclude with a few comments on this result: 

\begin{itemize}
 \item 
 The relation $I_c=I_{\rm QG} = 4\pi$, also  determines the relevant number of $e$-foldings in the inflationary epoch which carries the cosmic information. This is given by $N_e = (3\pi/2)I_c = 6\pi^2 \approx 59$, which  provides  an adequate amount of inflation. 

 \item Equation~(\ref{four}) can be inverted to express the cosmic information content $I_c$ in terms of the three densities. As I mentioned earlier, if we use the  values for $\rho_\Lambda$ and $\rho_{\rm eq}$ known from observations and take $\rho_{\rm inf} =( 10^{15}\ \text{GeV})^4$, we find that: 
\begin{equation}
I_c = \frac{1}{9\pi} \, \ln \left( \frac{4}{27} \frac{\rho_{\rm inf}^{3/2}}{\rho_\Lambda\,\rho_{\rm eq}^{1/2}}\right) \approx 4\pi \left( 1 \pm \mathcal{O} \left(10^{-3}\right)\right)
\label{six}
\end{equation} 
That is, the current observations  \textit{show that the CosmIn indeed has a value $4\pi$ to the precision of one part in a thousand!} Because of the logarithmic dependence on the cosmic parameters in \eq{six}, this result is  fairly stable and renders a purely observational support for the claim $I_c= I_{\rm QG} = 4\pi$.

\item Theoretically, one would like to determine the value of $\rho_{\Lambda}$ in terms of other parameters. Observationally, we can determine the values of $\rho_{\rm eq}$ and $\rho_\Lambda$ very well today but  have no direct handle on $\rho_{\rm inf}$. Using \eq{five}, we can  \textit{predict} the value of $\rho_{\rm inf}$ in terms of the well-determined parameters $\rho_{\rm eq}$ and $\rho_\Lambda$. We then get  $\rho_{\rm inf}^{1/4} = 1.2 \times 10^{15}$ GeV,  which is again a remarkable result.\footnote{In the calculation leading to \eq{six}, I assumed  that the reheating is instantaneous; ambiguities in the reheating dynamics can change this result by a factor of about 5, leading to the prediction $\rho_{\rm inf}^{1/4} \approx (1-5) \times 10^{15}$ GeV.} This is probably the \textit{only} model with quantum gravitational inputs \textit{which leads to a falsifiable prediction}. 

\end{itemize}

I believe the result obtained above is just a tip of the iceberg. It tells you that microscopic quantum gravitational physics can leave its trace in cosmic scales and, in fact, probably can be tested only in cosmology. Further exploration will require a systematic computation of corrections to \eq{evl1} and \eq{evl2}  in a context similar to non-equilibrium thermodynamics.

\section*{Acknowledgements} My research is partially supported by the J. C. Bose Research grant of DST, India. I thank 
Sumanta Chakraborty, Shyam Date, Tamara Davis, Dawood Kothawala,   Kinjalk Lochan and  Aseem Paranjape for comments on an earlier draft.

\end{document}